# Dynamics of water and solute transport in polymeric reverse osmosis membranes via molecular dynamics simulations

Meng Shen [a], Sinan Keten [a, b] and Richard M. Lueptow [a,*]

*Department of Mechanical Engineering, Northwestern University, 2145 Sheridan Road, Evanston, IL 60208-3111, USA*

*Department of Civil Engineering, Northwestern University, 2145 Sheridan Road, Evanston, IL 60208-3111, USA*

**Abstract:**

The Ångström-scale transport characteristics of water and six different solutes, methanol, ethanol, 2-propanol, urea, $Na^+$, and $Cl^-$, were studied for a polymeric reverse osmosis (RO) membrane, FT-30, using non-equilibrium molecular dynamics (NEMD) simulations. Results indicate that water transport increases with an increasing fraction of percolated free volume, or water-accessible open space, in the membrane polymer structure. The trajectories of solute molecules display Brownian motion and hop from pore to pore as they pass through the polymer chain structure of the membrane. The solute transport depends on both the Van der Waals size of the dehydrated solute and the electrostatic interaction of the solute with water and the membrane. For alcohol solutes, transport decreases for solutes with larger Van der Waals volume, which corresponds to less available percolated free volume, or solute-accessible space, within the membrane polymer structure. Urea has reduced transport compared ethanol, most likely due to more complex chemistry or polarity than the alcohol solutes, even though urea has a smaller Van der Waals volume than ethanol. $Na^+$ and $Cl^-$ experience the lowest transport, likely due to strong ion-water and ion-ion electrostatic interactions. NEMD simulations provide a unique opportunity to understand molecular level mechanisms for water and solute transport in polymeric RO membranes for water purification.

**1. Introduction:**

Advances in water purification technology have been fueled by an increasing scarcity of fresh water and a growing demand for water of high purity [1]. Membrane filtration has been adopted as a viable means for water purification since the 1960's [2], and polymeric reverse osmosis (RO) membranes are the dominant desalination technology today [3]. While RO membranes have been successful for desalination and water purification, the physico-chemical processes at the molecular scale are not well understood. Essentially, individual water molecules pass through the polymeric structure of the membrane, while, if all goes well, contaminants are somehow prevented from passing through the membrane. The reality is that polymeric RO membranes do not completely reject contaminants, and water permeability can always be improved. Although the rejection of large organic compounds, such as isoxathion, a pesticide [4], and ionic contaminants, such as sodium chloride, have been particularly successful, it is more challenging to achieve high rejection for small neutral organic compounds, such as methanol, ethanol, 2-propanol, and urea, which are potentially harmful for human health [5].

[*] Corresponding author: r-lueptow@northwestern.edu

Improvements in the reverse osmosis process are typically aimed at maximizing water flux and salt rejection, while minimizing fouling and energy consumption. Many advances have been made in the RO process in terms of module design, water pre-treatment, and energy recovery [6]. However, the greatest gains have come from the improvement of the RO membrane materials [3, 7]. Polymer based thin-film composite (TFC) membranes have achieved overwhelming success since the 1970's [3]. However, improvements in conventional polymeric membranes have been hindered since the 1990's [3] by the lack of a fundamental understanding of the physico-chemical processes in the membrane, especially at the molecular level. Continuum theories, such as the solution-diffusion model [8], the pore flow model [9] and the Nernst-Plank model [10] have been widely adopted to predict membrane performance and explain experimental results [8, 11]. However, these models are based on macro-scale assumptions for transport mechanisms, rather than a molecular level understanding of the RO membrane and its interaction with the solute and solvent.

Molecular dynamics (MD) simulations offer a powerful computational method to bridge the gap between the experimental observations and macroscopic theories for membrane filtration. MD simulations are based on explicit descriptions of atomic level details, namely, the atomic positions and inter-atomic force fields. This makes possible the study of membrane/water, membrane/solute, and water/solute interactions at the molecular scale. The use of simulations further permits comparisons of various solutes (organic/inorganic, charged/uncharged, large/small) through various polymeric membrane structures so that multiple factors governing transport and rejection mechanisms can be examined and compared. Furthermore, MD simulations make it relatively easy to extract atomic level dynamical data, such as solute trajectories and free volume distribution in the membrane [12], which are helpful for understanding the transport and rejection mechanisms.

Two types of MD have been used to study transport through membranes: Equilibrium MD (EMD) and Non-Equilibrium MD (NEMD). In EMD, no bias force is introduced, and transport properties are obtained from mean square displacements or from integrals of correlation functions based on the linear response theory [13]. On the other hand, in NEMD, non-equilibrium conditions such as a pressure difference [14] or concentration difference [15] across the membrane can be introduced so that transport phenomena can be directly observed at the Ångström scale. EMD has been used to examine transport behaviors of water molecules through carbon nanotubes [16, 17], while NEMD has been used to demonstrate the potential use of a zeolitic metal-organic framework (MOF) for water desalination [18]. In addition, NEMD has been used to study gas permeation through membranes [19, 20], although we note that gas permeation differs from reverse osmosis in terms of its driving force. The driving force for gas separation mainly results from partial pressure differences, essentially differences in gas concentrations, while the driving force for reverse osmosis comes from the hydraulic pressure difference across the membrane [8, 21, 22]. Thus, reverse osmosis can only be modeled as a diffusion process after expressing chemical potential in terms of the pressure difference at the membrane/permeate boundaries [8]. Furthermore, gas permeation is substantially different from reverse osmosis in that gases are sensitive to pressure and temperature changes, while water is incompressible. In some cases, EMD simulations have been used to study simplified pore structures, such as the effect of hydration on ion transport across idealized tubular reverse osmosis channels [23]. Although this simplified tubular model helped to reveal solute hydration as a key factor in ion rejection, simulations of complex atomic structures of actual RO membranes, like those described here, allow the analysis of other factors, such as the membrane structure and the solute/membrane chemistry, that are also important. According to a recent systematic review by Ebro et al. [24], while structurally simple inorganic membranes, particularly carbon nanotube membranes, have been studied to some extent by MD [15-17, 25-27], more commonly used polymeric membranes are rarely investigated using MD due to the difficulty in simulating complex polymeric RO membrane structures.

Here we consider an all-atom model of a common polymeric RO membrane known as FT-30, a crosslinked amide formed by the reaction between m-phenylene diamine (MPD) and trimesoyl chloride (TMC) monomers [28]. Since backbone rotations of the FT-30 polymer are prohibited due to crosslinking, the pore size is more restricted than rubbery polymers, making it useful for RO applications. Previous studies of the FT-30 membrane by EMD have made some progress [29-34]. Specifically, Kotelyanskii et al. observed a "jump" diffusion process for water [29] and a lower mobility of $Cl^-$ than $Na^+$ in the hydrated membrane by inserting $Cl^-$ and $Na^+$ ions into random locations in the membrane [30]; Harder et al. used EMD to determine water flux across the membrane [31]; Luo et al. used EMD to predict ion rejection from the position dependent free energy, ion pathways, and water flux based on a solubility-diffusion theory [32]; Hughes et al. used EMD and umbrella sampling methods to determine the free energy surface associated with selected ion pathways [33]; Kolev et al. generated the atomic models of a polyamide membrane that closely match the known characteristics of commercial membranes [34]. Nevertheless, the current understanding of the membranes can be pushed even further with NEMD simulation (as opposed to EMD simulations), because NEMD does not depend on any assumptions for the water permeation or the solute flow, and it allows the study of the mechanisms of the RO process under non-equilibrium conditions typical of practical RO membrane applications.

While previous studies have focused on ion transport across polymeric RO membranes [32, 33], the transport and rejection of small neutral organic molecules has not been studied at all using MD. Based on experiments, the transport and rejection of neutral molecules depends on solute sizes and molecular structures [5]. Although solute size can be taken into account approximately using steric models [11, 35], molecular-level mechanisms of solute rejection by size have not been investigated. Furthermore, the dependence of transport and rejection on solute structure and chemistry is not clearly understood, mostly because the atomic details, such as the solute shape, polarity, and coordination number in the hydrated state cannot be easily studied via experiments. Molecular dynamics offers the opportunity to investigate the molecular-level mechanisms of solvent and solute transport including atomic level details such as membrane molecular structure and chemistry, solute size, structure, and chemistry, and solute/solvent/membrane interactions.

In this paper, we use NEMD to explore the Ångström-scale transport of water, ions, and small organic solutes for a common polymeric membrane, FT-30. We begin by describing the methods for membrane molecular model construction and NEMD simulations. Then we relate the water transport to the membrane structure. Next we explore the dependence of the transport and rejection of organic solutes and ions on the membrane structure, solute structure, and solute hydration, and present the challenges for further research on RO membranes by MD.

## 2. Methods

### 2.1. Construction of the atomic models of the membrane

A crosslinked polymeric FT-30 membrane was computationally constructed using a heuristic method [31]. In this approach, TMC and MPD monomers move about randomly in a computational box. When the functional groups of the monomers that cross-link are within a specified distance from one another, a bond is inserted between the two monomers, forming an amide (Fig. 1(a)), eventually building up a polymeric structure (Fig. 1(b)). The computational crosslinking procedure starts from unreacted monomers rather than linear polymer chains in the same way that the actual crosslinked FT-30 membrane forms directly by reaction between TMC and MPD monomers [28]. Specifically, 192 TMC and 300 MPD monomers, the approximate stoichiometric ratio for the reactants, were first annealed at 1000 K and constant volume in a box of 55.1 Å $\times$ 53.8 Å $\times$ 34.2 Å for 1 ns with periodic boundary conditions. Next the system was relaxed at 340 K and constant volume with periodic boundary conditions in the x- and y- directions but a vacuum in the z- direction for 1 ns. Then the simulation was continued at 340 K but interrupted every 2 ps so that new amide bonds could be formed by deleting a chloride atom in the TMC acyl functional group and a hydrogen atom in the MPD amino functional group when a acyl carbon of a TMC monomer or residue was within a distance of 3.5 Å from an amino nitrogen of an MPD monomer or residue. Then a covalent bond was applied between the acyl carbon and the amino nitrogen. Energy minimization was performed after every crosslinking step. The crosslinking process slows down with time, because the diffusion of monomers is hindered as crosslinked clusters grow. To speed up the crosslinking process, after 2 ns, the distance criterion was increased to 4.5 Å. The structure after 4 ns of the virtual crosslinking simulations is shown in Fig. 1(b), where different colors represent different fragments, or chains, that are crosslinked. The largest chain in the membrane shown in Fig. 1(b) contains 411 monomers, the second largest chain contains 17 monomers, 10 smaller clusters contain 2 to 8 monomers, and 10 monomers remain unreacted. The unreacted monomers are left in the membrane so that the effects of residue monomers on flux can be included. The loose polymer surface structure is consistent with that found recently using novel transmission electron microscopy techniques [36]. To determine the isotropy of the membrane structure we calculated Herman's order parameter, defined as $S = \frac{1}{2}\left(3\langle\cos\theta\rangle - 1\right)$, where $\theta$ is the angle between the molecular symmetrical axis of the chain monomers and the z-direction [37]. $S$ can vary between 1 (fully ordered chains along the z-direction) and -1/2 (chains lying in the x-y plane), and $S$ is 0 for a fully isotropic structure. For the membrane shown in Fig. 1(b), $S$ along the z- direction fluctuates around 0 (Fig. S1 in the supplementary material), indicating an isotropic structure.

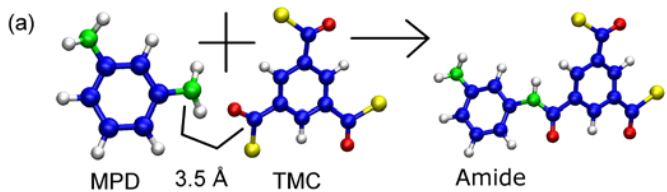

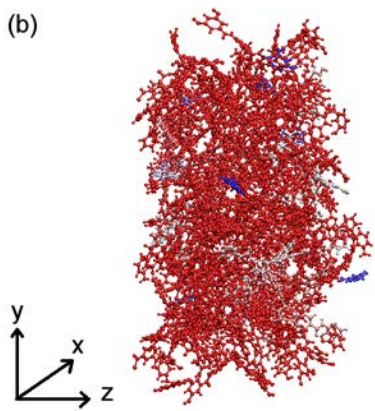

Fig. 1. (a) The monomers and the crosslinked amide, where green represents N, yellow represents Cl, blue represents C, white represents H, and red represents O, and (b) The membrane structure after the crosslinking process, where colors represent different crosslinked chains. (Color online.)

The Generalized AMBER Force Field (GAFF) [38] was used for constructing the TMC and MPD monomers and the FT-30 polymeric membrane [32]. The molecular dynamics simulations were performed by NAMD [39] with a time step of 1.0 fs. The SHAKE algorithm was used to constrain covalent bonds involving hydrogen atoms [40] to equilibrium values. The Particle-Mesh Ewald (PME) method was used to calculate the long-range electrostatic forces [41] with a grid spacing of 1.0 Å. The cross-linking operations were automated using a LINUX shell script. Four membranes, M1, M2, M3 and M4 were constructed using this approach.

**2.2. Molecular dynamics simulations of the reverse osmosis process**

The set-up for the NEMD simulations of water and solute transport is shown in Fig. 2. First the membrane was hydrated by filling in the open space with water by VMD, a visualization program for atomic structures [42]. The hydrated membrane was then placed between a reservoir of an aqueous solution with 192 solute molecules or ion pairs and 5000 water molecules on the left, corresponding to a solute concentration of about 2.13 M, and a reservoir of pure water with 5000 water molecules on the right. Graphene sheets created in XenoView [43] were added to the free surfaces of the solution reservoir on the left and pure water reservoir on the right. A pressure difference across the membrane was introduced by applying forces to each atom in the graphene sheets with the force on the left greater than that on the right. This directly simulates the pressure-driven operating conditions of RO membranes, which is quite different from previous simulations of water transport in concentration-driven forward osmosis [15]. The system was coupled to a global thermostat set at 300 K using Langevin dynamics.

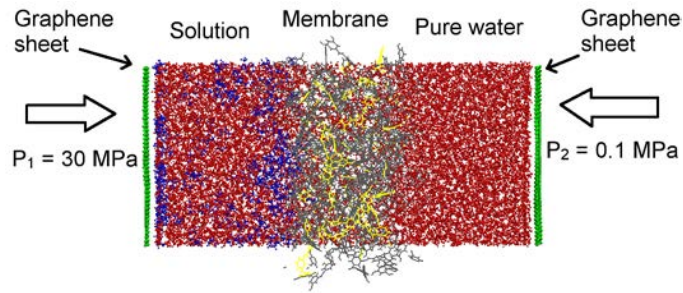

Fig. 2. Simulation setup for non-equilibrium molecular dynamics (NEMD) simulations with membrane M1. Water molecules are red, 2-propanol molecules in the solution to the left of the membrane are blue, graphene carbon atoms are green, and membrane atoms are gray except those that are pinned to a fixed position in space, which are yellow. (Color online).

Periodic boundary conditions were used in the depthwise x- and vertical y-directions with the computational domain extending approximately 55.1 Å in the x-direction and 53.8 Å in the y-direction. A vacuum was applied for a distance of 50 Å to the outside of each graphene sheet to avoid interactions between the atoms in the reservoirs across the boundary. The box sizes in the x-, y- and z-directions were kept constant. Unlike previous studies where there was only one reservoir [30, 32], this configuration allows us to mimic the experimental filtration process. To allow vibrations of most membrane atoms while fixing the membrane in the z direction, 10% of the membrane atoms were pinned to fixed points in space during the simulations (yellow polymer chains in Fig. 2). Fixing some of membrane atoms may slightly affect the macroscale water flux due to compression of the membrane at operating pressures, but it does not alter the Ångström-scale transport mechanisms within the membrane molecular structure [29], which is the focus of this study. Furthermore, fixing some membrane atoms can be viewed as similar to the typical thin-film composite RO membrane structure in which the thin polyamide RO film is supported on a porous polysulfone structure as well as by the laterally adjacent membrane structure.

Four organic solutes, methanol, ethanol, 2-propanol and urea, all of which are highly soluble in water, were studied and compared with sodium chloride. GAFF [38] was used as the force field for the organic solutes and membrane for the simulations. The parameters for urea were taken from Ref. [44]. The parameters for the other organic solutes were obtained from ANTECHAMBER 1.27 for the covalent bonds with the partial charge based on AM1-BCC charges [45, 46]. The parameters of the 6-12 Lennard-Jones potentials for $Na^+$ and $Cl^-$ were taken from Ref. [47]. The TIP3P potential was used for water [48]. The NEMD simulations were performed by NAMD [39] with a time step of 1.0 fs. The covalent bonds involving hydrogen atoms were constrained to equilibrium values by the SHAKE algorithm [40] with a cut-off for the non-bonded potential of 12.0 Å. The long-range electrostatic forces were calculated using the Particle-Mesh Ewald (PME) method [41] with a grid spacing of 1.0 Å. The switch distance for the Lennard-Jones potential was 10 Å.

Before the pressure-driven transport simulations commenced, forces corresponding to pressures of 0.1 MPa were added to the graphene planes on both sides, and the simulations were run for about 10 ns until water saturated the membrane. The thickness of the hydrated membrane M1 shown in Fig. 2 is 34.6 Å based on the concept of a Gibbs Dividing Surface [49]. The density profile of membrane M1 hydrated in pure water is shown in Fig. S2 in the supplementary materials. The thicknesses and densities of the other membranes are indicated in Table 1. The densities of hydrated membranes are less than dry membranes due to membrane swelling during hydration [50]. The membrane densities at 150 MPa, the highest pressure we studied, are slightly less than the densities at 1 atm, perhaps because the membrane becomes more fully hydrated at higher pressures in the short duration of the simulations. The membrane densities in the dry state are close to the experimental value of 1.24 g cm$^{-3}$ [50] and a previous simulation value of 1.20 g cm$^{-3}$ [34]. The membrane densities in the hydrated state are quite similar to the experimental result of 1.06 g cm$^{-3}$ [51] and previous simulation results of about 1.10 g cm$^{-3}$ [32, 33]. The membranes under study are much thinner than actual polymeric membranes used commercially, which are typically about 0.2 µm thick [7]. As a result, these simulated membranes would not be expected to exactly reproduce the macroscale water flux or solute rejection data from experimental studies. Nevertheless, we show later that the simulated water flux is in the middle of the range expected for commercial membranes. More importantly, the atomic level structure of the simulated membrane is adequate to provide an opportunity to understand the effects of the membrane's Ångström-scale molecular structure on water and solute transport.

Table 1. The density and thickness of dry and hydrated membranes.

| Membranes | Dry density (g cm$^{-3}$) | Dry thickness (Å) | Hydrated density at 1 atm (g cm$^{-3}$) | Hydrated thickness at 1 atm (Å) | Hydrated density at 150 MPa (g cm$^{-3}$) | Hydrated thickness at 150 MPa (Å) |
|---|---|---|---|---|---|---|
| M1 | 1.20 | 30.6 | 1.07 | 34.3 | 1.06 | 34.6 |
| M2 | 1.17 | 31.6 | 1.08 | 34.2 | 1.05 | 34.9 |
| M3 | 1.23 | 29.8 | 1.12 | 32.7 | 1.10 | 33.2 |
| M4 | 1.23 | 29.8 | 1.10 | 33.3 | 1.08 | 33.8 |

In the NEMD simulations of pressure driven transport, pressures on the solution side were varied from 30 MPa to 150 MPa, while the pressure on the pure water side was maintained at 0.1 MPa. Transport of water and solutes was analyzed using scripts written in TCL, a scripting language adapted to VMD, in which the number of water/solute molecules within the membrane and on both sides of the membrane were determined every 0.03 ns.

## 3. Results and discussion

## 3.1. The pressure dependence of water transport

Fig. 3 shows pure water transport through membrane M1 at 30, 60, 90, 120 and 150 MPa, in terms of the number of water molecules passing through the membrane as a function of time. Note that the extremely high trans-membrane pressures, more than an order of magnitude higher than that normally used in desalination applications, were necessary to assure a statistically meaningful number of molecules passing through the membrane in the very short time that was feasible to simulate. As expected, after about 10 ns, the water molecule flux (slope of the curve) is approximately constant and increases with increasing pressure. The variation in the water flux during the initial 10 ns occurs because some unreacted single monomers close to the membrane surface are gradually driven out of the membrane M1 during that period of time. Similar results occur for the other membranes.

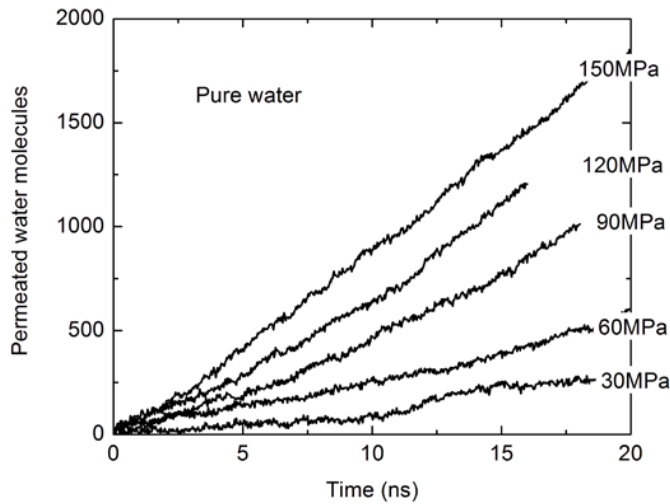

Fig. 3. The number of water molecules transported through the membrane as a function of time for pure water at various pressures for membrane M1.

The water molecule flux for pure water and several contaminant solutions, in terms of the number of water molecules passing through the membrane per unit area per unit time and shown in Fig. 4 as a function of pressure, was calculated from the cross-sectional area of the membrane and the slope of the linear regime of each curve in Fig. 3. Ideally, the pure water flux-pressure curve should be linear and intersect the vertical axis at the origin because the osmotic pressure is zero. The flux curve in the simulations is not perfectly linear, but quite close, perhaps because a few unreacted TMC or MPD monomers or residues remain in the membrane at low pressures, but are driven out of the membrane at high pressures.

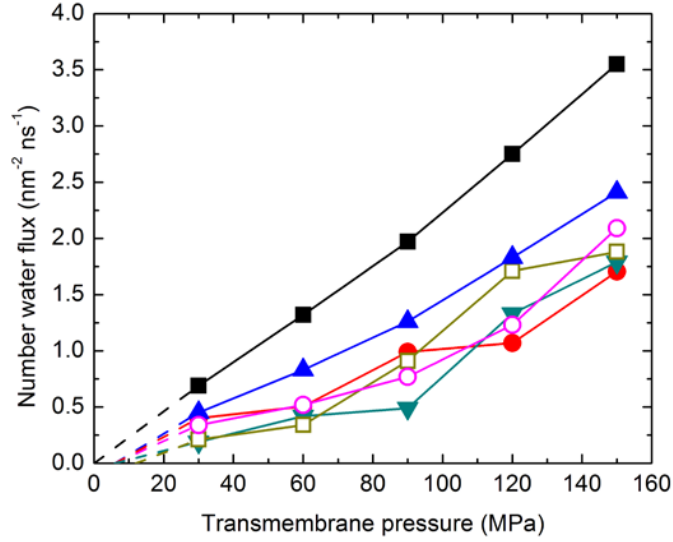

Fig. 4. Water molecule flux as a function of pressure for (■) pure water, (○) methanol, (▲) ethanol, (●) 2-propanol, (▼) urea, and (□) NaCl solutions on the left side of the membrane. Dashed lines represent extrapolations based on the theoretical osmotic pressure. (Color online.)

For comparison, the experimentally-measured macroscale water flux for a typical commercial RO membrane is in the range of $2.4 \times 10^{-5}$ to $9.8 \times 10^{-5}$ m s$^{-1}$ at 4.1 MPa, the maximum recommended operating pressure for several typical membranes [5]. Recall that the water flux can be written as:

$$J_w = -\frac{K(\Delta P - \Delta \pi)}{L} \quad (1)$$

where $\frac{K}{L}$ is the permeability, $K$ is the permeability coefficient, $L$ is the membrane thickness, $\Delta P$ is the pressure difference applied across the membrane, and $\Delta \pi$ is the osmotic pressure. Thus, the water flux is inversely proportional to the membrane thickness and is directly proportional to the pressure difference applied across the membrane after accounting for the osmotic pressure. Based on this relation, the water flux for the 35 Å thick membrane operated at 150 MPa in this study should proportionally be in the range of 0.052 to 0.21 m s$^{-1}$. This macroscale flux range is equivalent to a water molecule number flux of 1.7 to 6.9 nm$^{-2}$ ns$^{-1}$. The pure water molecule number flux of 3.55 nm$^{-2}$ ns$^{-1}$ at 150 MPa in Fig. 4 is in the middle of this range. Similar comparisons with other experimental values for the macroscale water flux [7, 52] indicate values for the water molecule flux of 1.30 to 6.00 nm$^{-2}$ ns$^{-1}$ adjusted to the same conditions as this study, again consistent with the simulated water molecule flux here. Thus, the results in Fig. 4 for the pure water molecule flux are consistent with macroscale experimental results when properly accounting for the reduced membrane thickness and higher operating pressure that are necessary for reasonable computational times for NEMD simulations.

The results for the water molecule flux for the various 2.13 M solutions at various transmembrane pressures are also indicated in Fig. 4. These simulation results are based on the number of permeated water molecules as a function of time for pure water and contaminant solutions, an example of which is shown for 150 MPa in Fig. 5. In all cases, the water molecule flux with contaminants is less than the pure water flux, likely due to combined effects of concentration polarization and osmotic pressure. Returning to Fig. 4, the reduced water flux with contaminants present is reflected at all transmembrane pressures. Furthermore, at low pressures the data is consistent with what is expected when extrapolating the curves to the osmotic pressure for 2.13 M solutions: about 12 MPa for NaCl [53, 54] and about 6 MPa for the organic solutes.

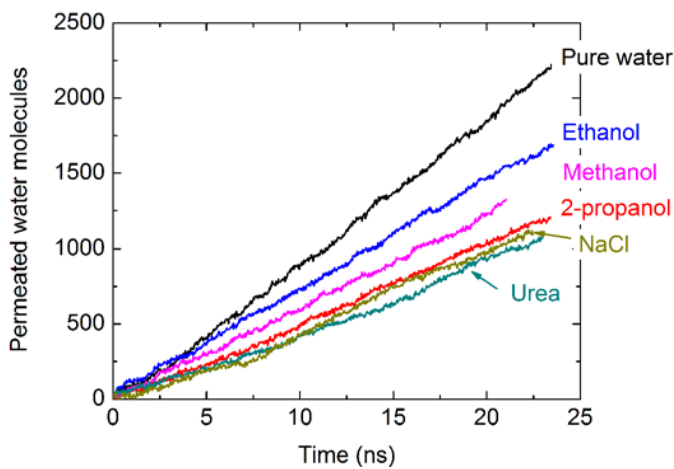

Fig. 5. The number of water molecules transported through the membrane as a function of time in membrane M1 for pure water and methanol, ethanol, 2-propanol, urea, and NaCl solutions at 150 MPa. (Color online.)

### 3.2. Membrane structure and its effect on water transport

To study the effect of the membrane molecular structure on water transport, we consider all four virtual membranes, M1, M2, M3, and M4, each of which has a different structure at the molecular level due to random variations in the initial configuration during the synthetic polymerization process. Since the mass density of all four membranes is similar, the differences in the membrane pore structure, which can be analyzed in terms of the size and connectivity of the water-accessible space, or free volume, in the membrane structure at the molecular level, likely lead to different degrees of water transport. To show this, the membrane pore structure was analyzed by considering the connected volume accessible by a probe that is 1.4 Å in radius, which corresponds to the mean Van der Waals radius of water [55], using a modified version of the Poreblazer code [56], which is based on the Hoshen-Kopelman cluster labeling algorithm [57]. The free volumes of the four membranes in the hydrated state are displayed in Fig. 6. The colored area indicates the percolated open spaces in the membrane structure that are wide enough for water molecules to pass through, with colors representing each of 111 different adjacent 0.5 Å thick y-z planes. The modified Poreblazer code identifies only pores through which a water molecule (based on the 1.4 Å probe radius) can pass from one side of the membrane to the other. Thus, only the percolated volume through which water molecules can pass through the entire membrane is shown in Fig. 6. We clarify here that the terminology "pore" used here means the water-accessible space within the membrane polymer structure and is not meant to describe or imply cylindrical or tubular pores characteristic of a pore flow model.

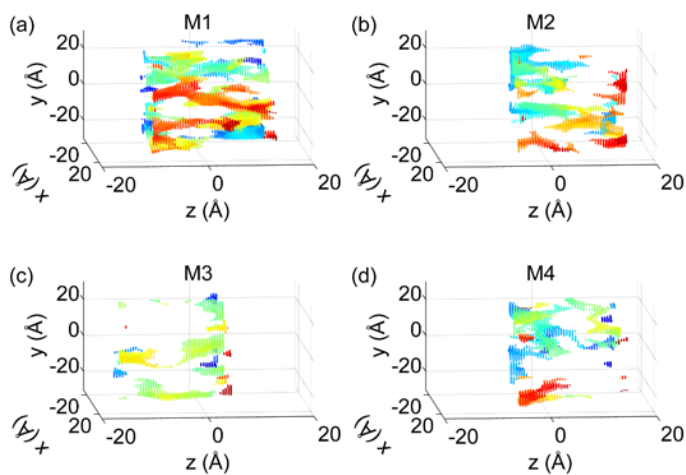

Fig. 6. The free volume distribution for hydrated membranes (a) M1, (b) M2, (c) M3 and (d) M4. Each color represents a 0.5 Å thick plane at a different depth in the x direction. (Color online.)

The differences in the total percolated free volume between the four membranes are striking, considering that all membranes have similar mass densities (Table 1). It is worth noting that the four membranes have the same degree of crosslinking; namely, 85.6 % of the functional groups in TMC were cross-linked. Therefore, the differences in the percolated free volume are not a result of different degrees of crosslinking, but due to different spatial distributions of the monomers and crosslinking sites resulting in different degrees of space between polymer chains at the molecular level. The performance of the membrane is most easily considered in terms of the pure water permeability coefficient, $K$, from equation (1). The percolated free volume directly correlates with $K$, as shown in Fig. 7, indicating the importance of connectivity of the free volume for water transport. Only percolated pores facilitate water transport, while dead-end pores can be occupied by water molecules but do not contribute to water transport.

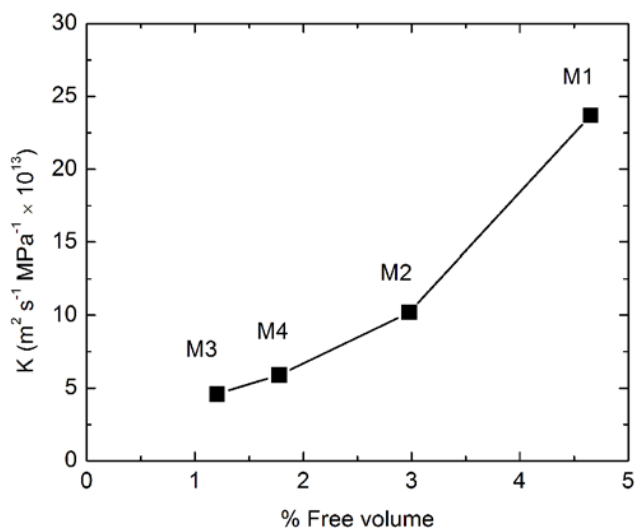

Fig. 7. The permeability coefficient $K$ (solid squares) as a function of the percolated free volume percentage in the hydrated state for four different membranes.

To further correlate the percolated free volume with water transport, we compare the trajectories of water molecules that traverse through membrane M1 with the percolated free volume within the membrane. Since the membrane is in constant motion due to thermal vibrations and collisions with the water molecules, the fluctuations across the entire membrane structure lead to dynamic pore dimensions. These membrane vibrations are surprisingly strong. A video of the membrane at 150 MPa viewed as shown in Fig. 2 over 50.7 ns is available as Supplementary Material (V1) to more clearly demonstrate the nature and magnitude of the membrane vibrations.

To take the dynamics pore dimensions into account, we consider the free volume accumulated over 10 ns during simulations in the hydrated state, a duration long enough for individual water molecules to travel through the entire membrane thickness. The size and connectivity of the accumulated free volume was analyzed using the modified Poreblazer code. For comparison, trajectories of water molecules that traverse the membrane from one side to the other were calculated based on the positions of the water molecules in the membrane every 0.03 ns during the same 10 ns portion of the simulation in the hydrated state. The accumulated free volume percolated in the z direction (Fig. 8(a)) matches well with the trajectories of water molecules passing through the membrane (Fig. 8(b)). Note that if the percolated free volume is calculated at a single instant, rather than accumulated over some finite time to account for dynamic pore dimensions, the percolated free volume is substantially less and does not correlate well with the trajectories of water molecules that traverse the membrane.

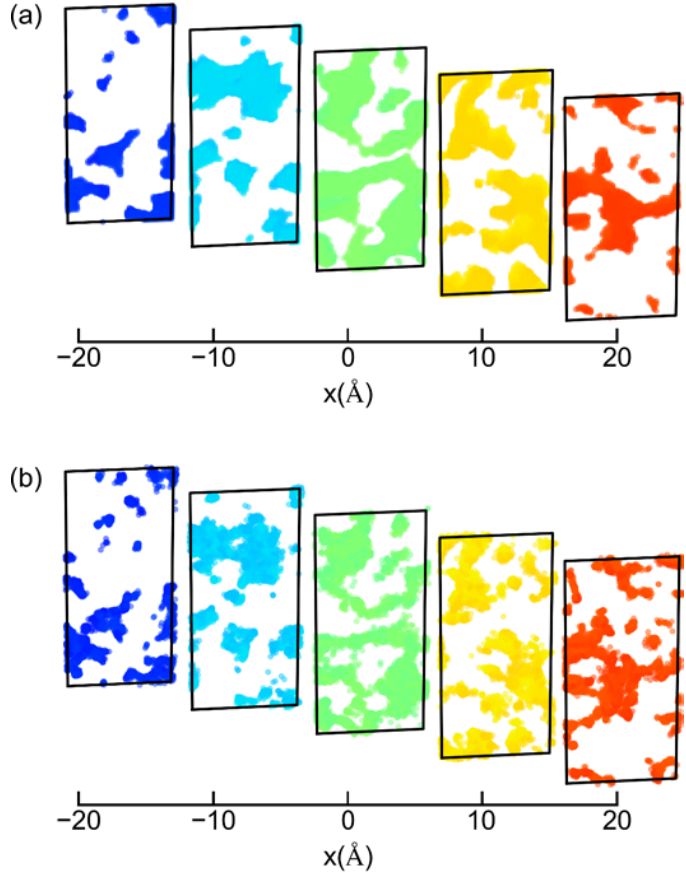

Fig. 8. (a) The percolated water-accessible free volume accumulated over 10 ns in the dense region of the membrane M1 and (b) the trajectories of water molecules that passed through the membrane M1 over 10 ns for 2 Å thick slices at x = -20 Å, -10 Å, 0 Å, 10 Å and 20 Å. The color represents the depth in the x direction. (Color online.)

Extensive crosslinking in FT-30 membranes results in permanent local voids [7, 34] that are evident in the dry state. However, it seems that these voids are enlarged in the static hydrated state due to local configuration changes in polymer segments in response to the presence of water solvent molecules, and they are further augmented at high pressures due to thermal vibration and collisions with water molecules leading to the increased free volume in the membrane. To confirm this, we compare the percolated free volume in the dry state (Fig. 9(a)) with that in the hydrated state (Fig. 9(b)) for membrane M1. The percolated pores in the dry state remain percolated in the hydrated state, but more pores are open in the hydrated state than in the dry state due to membrane swelling during hydration. Furthermore, the percolated free volume accumulated over 10 ns in the hydrated state (Fig. 9(c)), is about 4 times larger than the percolated free volume at a single instant in the hydrated state (Fig. 9(b)), as shown in Fig. 9(d), indicating that the vibration of membrane atoms related to thermal vibrations and collisions with water molecules has a significant effect on the free volume in polymeric RO membranes. This might be one of the major differences between polymeric RO membranes and more rigid zeolitic membranes [58]. The flexibility of the soft bonds (dihedral, Van der Waals and electrostatic bonds) in the polymeric RO membranes allows many "gates" to be transiently opened for water transport via the "dynamic membrane structure", perhaps explaining the higher water flux through polymeric membranes than through zeolitic membranes [59].

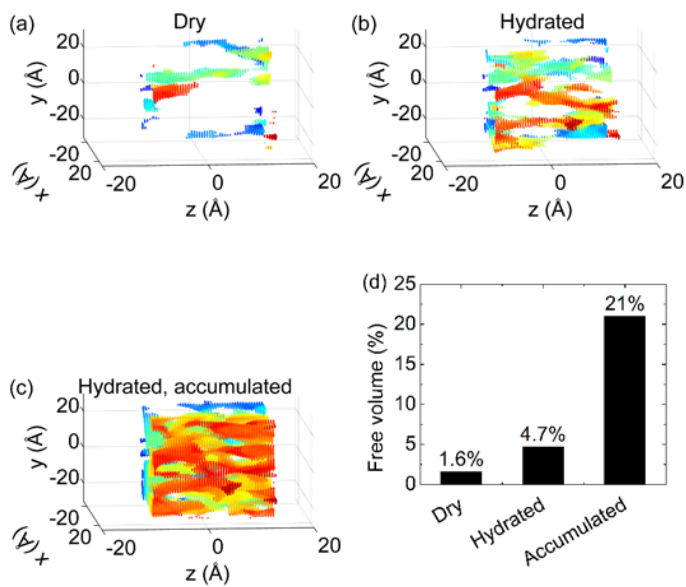

Fig. 9. The free volume for membrane M1 (a) in the dry state, (b) at a single frame in the hydrated state, and (c) accumulated over 10 ns in the hydrated state. (d) The percentage of free volume in the dense membrane region for the dry membrane M1, hydrated membrane M1 at a single frame and hydrated membrane accumulated over 10 ns. (Color online.)

The effective pore diameter of RO membranes has been estimated to be about 7 Å in previous experimental studies [5]. However, MD allows the determination of how pores sizes are distributed. The Pore Size Distribution (PSD) was evaluated in Poreblazer by a geometrical method where 10,000 points are randomly seeded in the percolated open regions of the membrane, and the radius of the largest sphere enclosing each point but not within a distance of the probe radius (1.4 Å, corresponding to the mean Van der Waals radius of water [55]) from the surfaces of membrane atoms is found [56]. Fig. 10 shows the PSD of the dry and hydrated membrane M1. The peaks in the distribution are likely a consequence of the limited data available and resulting statistical limitations due to the small membrane volume that could be simulated. The PSD for the accumulated percolated free volume in the hydrated state is slightly different from PSD of the dry membrane; in the hydrated state, the small pores are expanded, leading to more intermediate size pores, and consequently the largest pores make up a smaller percentage of the free volume. The highest peak of PSD of the hydrated membrane M1 is located at around 7 Å, consistent with the estimate from experimental rejection data [5]. The dry membrane peaks may correspond to the "network pores" and "aggregate pores" experimentally identified by positron annihilation spectroscopy [7]. Network pores are small spaces within aggregates formed by polymer segments, typically about 4.2-4.8 Å in diameter; while aggregate pores are large open spaces between polymer aggregates, typically about 7.0-9.0 Å in diameter [7]. To be sure that the overall pore size distribution is not affected by surface structure together with a very thin membrane, we have constructed a computational model of a membrane that is twice as thick as the one in Fig. 1b. The pore size distribution of the thicker membrane matches that of the thinner membranes we have used. It is worth noting that the PSD analysis used for these simulations does not suffer from membrane contraction after drying, which is inherent in PALS measurements.

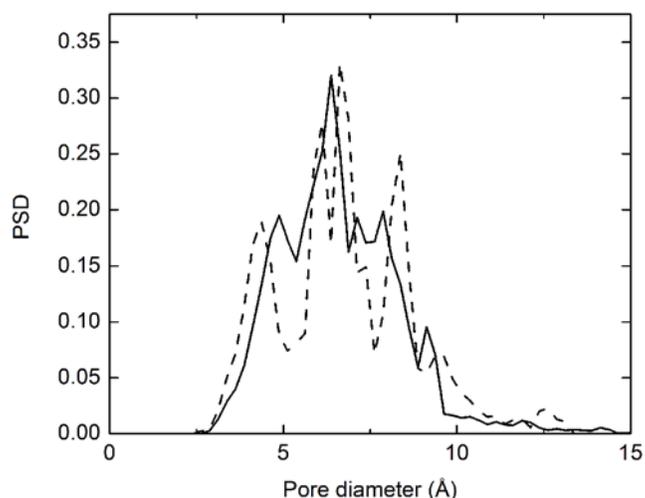

Fig. 10. Pore size distribution for the dry membrane M1 after the construction processes (dashed curve) and for the summation of over 10 ns in hydrated states in NEMD simulations (solid curve).

### 3.3. Solute transport

Our interest here is in the transport of solutes, particularly small organic solutes, within the membrane. The advantage of NEMD simulations is that we can track individual solute molecules as they traverse the membrane to get a physical sense of the transport at the molecular level. We note that the concentration of the feed solution varies by up to 38% during the NEMD simulations, much like the feed solution concentration varies in a typical stirred cell experiment. However, the solute transport is nearly independent of the solute concentration and is driven primarily by the pressure difference (see Fig. S3 in the supplementary materials). The trajectories of representative solute molecules in the membrane are shown in Fig. 11. The points connected by line segments in Fig. 11(a-h) represent solute locations that are 0.03 ns apart. In each case, the solute starts in the solution to the left of the membrane surface (indicated by vertical dashed lines) and passes from left to right through the membrane to the permeate side. Two views are shown: a projection of the trajectory in the x-z plane and a projection of the trajectory in the y-z plane. Note that in some cases the solute trajectory appears to have two separate portions, but this is merely the result of the solute passing across the periodic boundary of the computational domain.

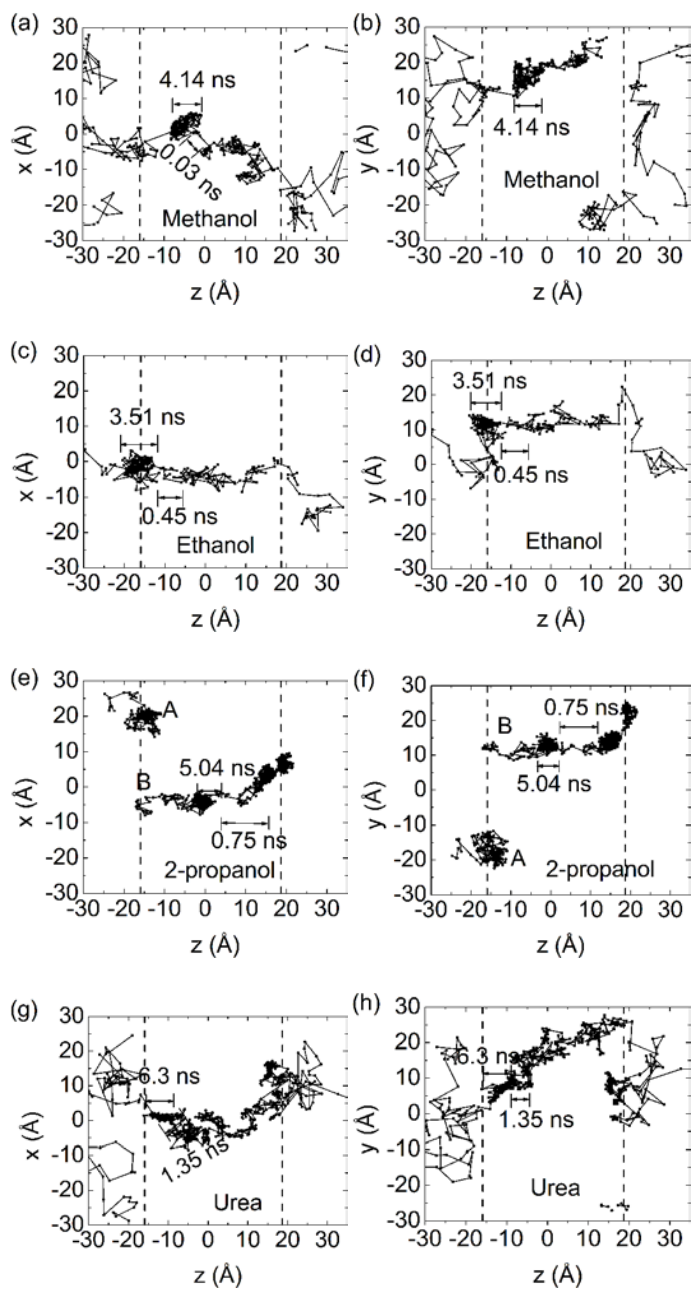

Fig. 11. Trajectories at a transmembrane pressure of 150 MPa for (a, b) methanol, (c, d) ethanol, (e, f) 2-propanol and (g, h) urea molecules illustrating "hopping" mechanisms, especially clear for 2-propanol. Residence time is indicated for selected pores and paths between pores. The approximate location of the membrane surface is indicated by the vertical dashed lines.

Consider first the trajectory of a methanol molecule in the M1 membrane in Fig. 11(a,b). The molecule initially follows a random path, Brownian motion, in the feed solution to the left of the membrane as it is buffeted by water and solute molecules. The molecule nears the membrane surface (evident in Fig. 11(b)) a couple of times before it actually enters the membrane. Once it enters, it travels relatively quickly in some parts of the membrane (where the interconnecting line segments between dots are sparse) and appears to become trapped in other parts, presumably in voids or pores. For instance, the methanol molecule becomes trapped in a pore for just over 4 ns, as indicated in Fig. 11(a,b), where it follows a random path that "fills" the pore as it collides with water molecules and polymeric chains of the membrane but remains confined by the membrane molecules bounding the pore. It is evident that the size of this pore is a little under 10 Å in diameter based on the extent of the confined random path of the methanol molecule. After about 4 ns in this pore, the methanol molecule "hops" very quickly, taking less than 0.1 ns, to another pore. This pore appears smaller, a little less than 5 Å wide, but elongated (12-15 Å long). The jump to a pore located at the coordinate z = 10 Å, evident in Fig. 11(b), is a consequence of the molecule crossing the periodic boundary near the top of the figure and re-entering at the bottom of the figure. From this pore, the methanol molecule eventually hops out of the membrane to the permeate on the right side of the membrane.

Similar results occur for the other solute molecules, but details depend on the specific trajectory, which is partly a consequence of the solute molecule size, as will be shown later. For instance, the ethanol molecule in Fig. 11(c,d) remains trapped in a pore near the solute side of the membrane for 3.5 ns before the random fluctuations bring it to a path where it quickly (0.45 ns) passes to an enlarged pore deeper in the membrane. Although it appears that the initial pore extends to the left of the membrane surface plane, this is not actually the case, because the membrane surface is just a loose tangle of the membrane polymer chains at the molecular level, as shown in Fig. 1(b) and Fig. 2, not a smooth, well-defined surface. The pore on the solute side of the membrane is in this loose tangle of polymer chains.

The trajectory for a 2-propanol molecule in Fig. 11(e,f) is somewhat different than those for the methanol or ethanol molecules. In particular, the volumes of the pores are smaller, about 5 Å, and jumps between pores follow a much narrower trajectory. This is likely a consequence of the larger size of the 2-propanol molecule. Because of its larger size and the resulting steric effects, its random motion within a pore is more limited than that for smaller molecules, and its side-to-side motion through the narrow connections between pores is more constrained. Nevertheless, the random motions of the 2-propanol molecule as well as the vibrations of the membrane structure are key to solute transport through the membrane. This is further amplified in a video of the 2-propanol trajectory over 38.9 ns (see Supplementary Material, V2). The video clearly demonstrates the role of Brownian motion and the resulting randomness in the solute molecule's path. With reference to Fig. 11(e,f) and the video, Brownian motion buffets molecules in the solution between the left graphene sheet and the membrane surface. The 2-propanol molecule first enters the membrane at A, but, in spite of the applied pressure difference, the molecule actually comes back out of the membrane. It eventually makes its way to another part of the membrane at B, where it finally makes its way through the membrane to its right side. The video of the molecule's motion also demonstrates how, once within the membrane, the molecule travels relatively quickly through certain portions of the membrane (where the dots and interconnecting line segments are sparse), while it remains temporarily trapped in voids, or "pores", at other locations.

In spite of having nearly the same molecular weight as 2-propanol (see Table 2), urea has a much different trajectory through the membrane. In fact, the random oscillations along its trajectory result in a much more widely spread path than any of the alcohol solutes, even the much smaller methanol molecule. This is likely a result of the more complex chemistry of urea and the polyamide membrane, which is currently under study.

Table 2. The side and head-on views of space-filling models of four organic solutes and a water molecule drawn by VMD [42]. The area of the head-on view shows the minimum cross-section area.

| Solute | MW (g mol$^{-1}$) | Side view | Head-on view |
| --- | --- | --- | --- |
| Water | 18.02 | 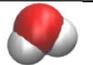 | 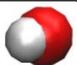 |
| Methanol | 32.04 | 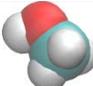 | 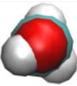 |
| Ethanol | 46.07 | 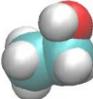 | 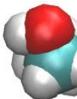 |
| 2-Propanol | 60.10 | 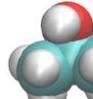 | 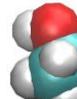 |
| Urea | 60.06 | 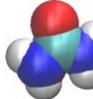 | 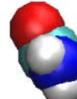 |

One of the most intriguing findings from the simulations is that the membrane itself is in constant motion due to thermal fluctuations and collisions with water and solute molecules, resulting in vibrations across the entire membrane's molecular structure, leading to dynamic pore dimensions [60]. As noted earlier, this is most clearly evident in a video corresponding to the orientation shown in Fig. 2 in which the trajectories of several 2-propanol molecules through the membrane are recorded (see Supplementary Material, V1). In this video, only the membrane molecules and a few 2-propanol molecules that make it through or close to the membrane are shown; no water molecules are shown. The red 2-propanol molecule in the video that is labeled with red lettering is the same molecule and viewed from the same angle as the trajectory shown in Fig. 11(f). Although the membrane is computationally pinned to several fixed points in space to keep it from translating, the degree of motion of the membrane in the video is remarkable as water and solute molecules collide with membrane polymer chains. Early in the video, the labeled 2-propanol molecule enters the lower left part of the membrane, but due to Brownian motion it exits back into the feed solution. It later enters the membrane near the upper left and passes through to the right side of the membrane, though it is held up in pores, similar to what is shown in Fig. 11(f). As noted earlier, the free volume for a hydrated membrane that is calculated without considering the membrane mobility at the molecular scale (Fig. 9(b)) is much smaller than the water-accessible space accounting for the dynamic membrane structure (Fig. 8 and Fig. 9(c)). This result, which would not be possible except from these MD simulations, makes clear that the local fluctuations in the solute trajectories and the dynamics of the polymer chains are important factors in the transport of both water and solutes.

We detour here briefly to consider concentration polarization, which is somewhat difficult to accurately model in these simulations. Fig. 12 shows the solute number density profile of the four organic solutes and NaCl along the z- coordinate before the transmembrane pressure was applied, and 20 ns after a 150 MPa pressure was applied to the graphene sheet to the left of the solution reservoir. The membrane itself spans approximately -20 Å < z < 20 Å. The high concentration of organic solute molecules at the far left in Fig. 12(a-h) is a result of the preferred orientation of the amphiphilic molecules at the water-graphene interface [61, 62]. This does not occur for NaCl because of the ionic nature of the solute. We note that the high solute concentration at the graphene sheet alters the effective solute concentration on left side of the membrane. However, for all of the organic solute density profiles, the high solute concentration polarization layer is still evident on the left surface of the membrane, and the high solute concentration extends into the membrane, likely as a result of diffusion due to the concentration gradient from one side of the membrane to the other. This is even evident before the transmembrane pressure is applied. In fact, the concentration polarization decreases to some extent after the pressure is applied, most evident for 2-propanol in Fig. 12(e,f). This likely comes about due to depletion of solute molecules on the left side as solute molecules pass into the membrane. The decrease of concentration polarization is the least evident for methanol in Fig. 12(a,b), likely because it easily permeates into the membrane even before the reverse osmosis pressure is applied (Fig. 12(a)), and it quickly passes through the membrane after the transmembrane pressure is applied (Fig. 12(b)) due to its small size. Similarly, ethanol penetrates deep into the membrane even before the reverse osmosis pressure is applied (Fig. 12(c)), but passes through the membrane less quickly than methanol after the transmembrane pressure is applied (Fig. 12(d)), consistent with its larger size than methanol. Urea behaves differently than the alcohols. In spite of its smaller size than ethanol, it is trapped closer to the solution-membrane interface before the transmembrane pressure is applied (Fig. 12 (g)), and it passes through the membrane more slowly after the transmembrane pressure is applied, noting that no urea molecules are to the right of the membrane after 20 ns (Fig. 12 (h)). We attribute this to the propensity for urea to form hydrogen bonds with local water and polymer molecules more frequently than the alcohols, but this is still under study. The results are quite different for NaCl due to its ionic nature (Fig. 12 (i,j)). No concentration polarization is evident and very few ions even make it into the left surface of the membrane for the time scales that could be simulated here.

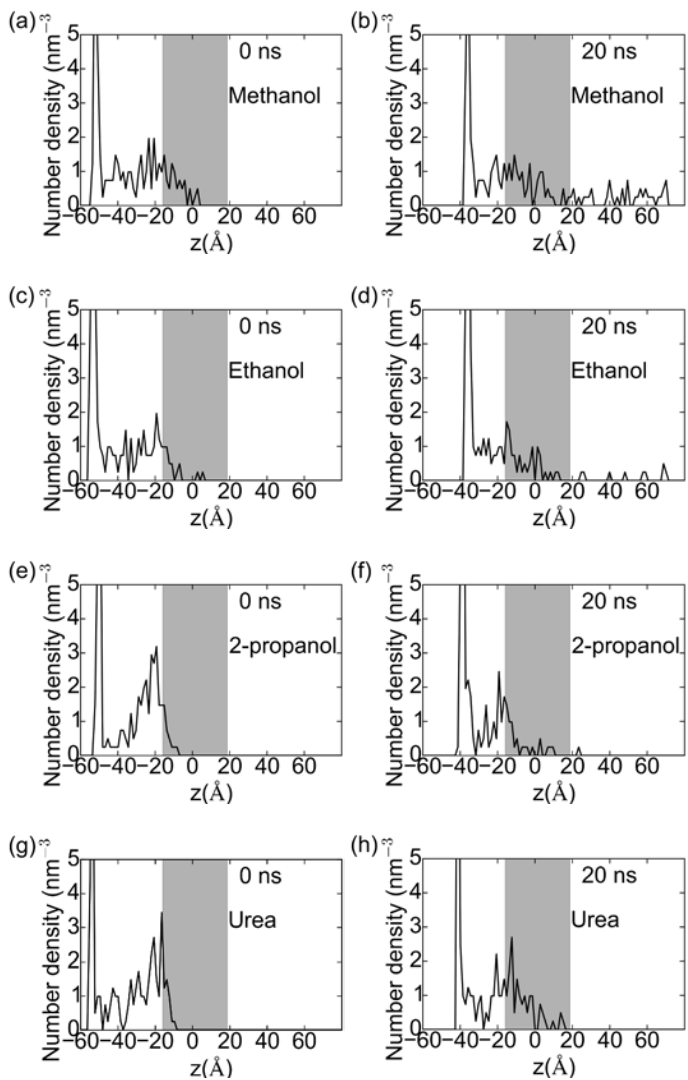

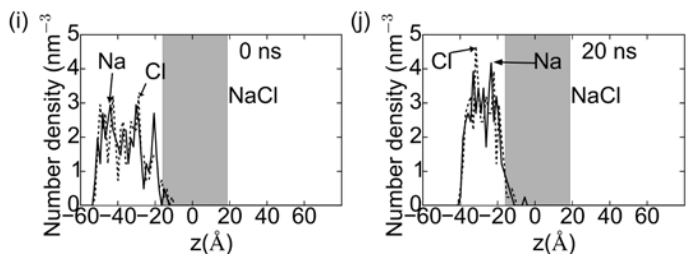

Fig. 12. The solute number density (a,c,e,g,i) before applying a reverse osmotic pressure and (b,d,f,h,j) 20 ns after applying 150 MPa pressure to the graphene layer on the left. The gray region represents the membrane.

For all of the organic solutes, some solute molecules permeate into the membrane, but the number density within the membrane is much higher for methanol and ethanol than 2-propanol—clearly a solute size effect. And it is evident that more small alcohol solute molecules (methanol and ethanol) pass through the membrane than larger 2-propanol molecules. Thus, the results in Fig. 12 clearly demonstrate how MD simulations offer insight into concentration polarization and rejection. However, these solute number density profiles also illustrate one of the challenges with MD simulations. The computational requirements for such simulations are immense, so it is possible to only consider a small number of solute and water molecules (192 and 5000, respectively, on the left side of the membrane). This limitation, along with the small size of the membrane that can be simulated, makes the quantitative study of concentration polarization difficult. Hence, we focus on molecular details of the transport of solutes through the membrane and limit the discussion of concentration polarization to the above qualitative results only. We further note that in spite of the somewhat different solute concentrations available on the feed side of the membrane (as a consequence of organic molecules accumulating to the graphene sheet and depletion of solute molecules as they pass through the membrane), there is always an adequate supply of solute molecules at the surface of the membrane, making possible the molecular level study of solute transport within the membrane itself.

In order to investigate the dependence of the solute transport and rejection on the solute size and structure, the transport of the four organic solutes (methanol, ethanol, 2-propanol and urea) and NaCl, are considered. Due to limitations in the simulation duration and domain size, it is difficult to accurately measure the macroscale rejection as would be done experimentally. Instead, we quantify the solute transport and rejection in the simulation by simply counting number of solute molecules (N) that pass through the central plane of the membrane by the time that 1000 water molecules have permeated through the membrane. The higher the permeation rate of the solute, the lower the rejection of that solute. A similar alternative definition for solute rejection was adopted in a previous simulation study [63].

The permeation rates of different organic solutes are in the order: methanol > ethanol > urea > 2-propanol at 150 MPa, as shown in Fig. 13. Although not readily evident in Fig. 13 due to the scale of the vertical axis, the permeation rate of urea is about 67% larger than that of 2-propanol. This is consistent with previous experimental rejection measurements [5]. During the NEMD simulations, Na$^+$ and Cl$^-$ ions hardly penetrate into the membrane, and none at all reach the central plane of the membrane (see Fig. 12(j)), so NaCl is not included in Fig. 13. The NaCl result confirms that the virtual membrane constructed in our simulation indeed performs like a typical reverse osmosis (RO) membrane, rejecting mono-valent salts.

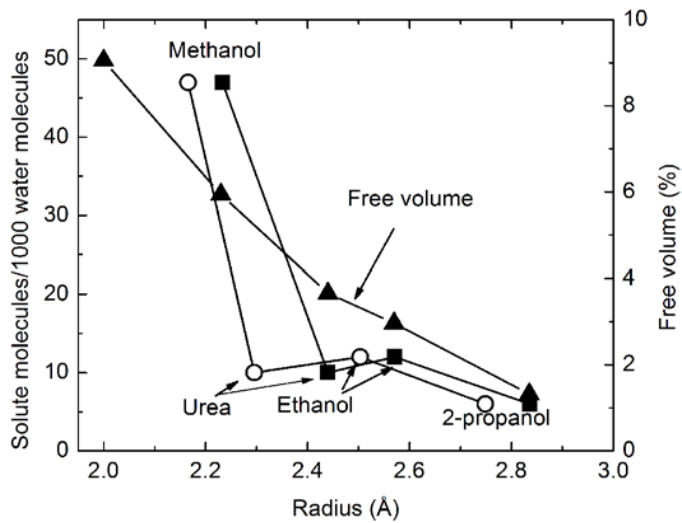

Fig. 13. The number of solute molecules passing through the central plane of the M1 membrane when 1000 water molecules have permeated through the membrane as a function of solute radius estimated from the solute volumes assuming a spherical shape (solid squares) and estimated from the minimum cross-section of the solutes assuming a circular cross-section (open circles). Both the solute volume and area are calculated from a grid-counting algorithm. The solute-accessible volume, or free volume, within the membrane is a function of the probe radius (solid triangles).

What is perhaps more important is that the solute permeability (measured in terms of the number of solute molecules per 1000 water molecules) for organic solutes roughly follows the solute-accessible volume calculated as described previously but based on varying probe radii. The solute-accessible volume represents the available space for a solute molecule to pass between polymer chains of the membrane, but such that it depends on the radius of the spherical probe, which is varied from 2.0 to 2.83 Å. Over this probe size range, the solute-accessible volume decreases from about 9 % down to only 1%, as shown in Fig. 13. To compare this to the size of the organic solutes, we characterized the volume of the solutes in two ways. First, the radii of the atoms were estimated using a 3D grid counting technique [64]. The solute volumes are in the following order: methanol < urea < ethanol < 2-propanol. The solute volumes are then converted into equivalent solute radii assuming a spherical solute shape. (Solute radii calculated here are slightly different from those used in Ref. [5], which are based on the Einstein Diffusion Equation. The grid counting method solely accounts for the static size of dehydrated solutes, while the Einstein Diffusion Equation considers the dynamics of a solute molecule in an infinitely dilute solution, which is not the case here.) The problem with assuming a spherical solute shape is that the molecules are actually quite asymmetric, as shown in Table 2. To account for this, we also determine the equivalent circular radius based on the minimum projected area (labeled "Head-on View" in Table 2). In this case a 2D grid counting method was used and converted to an equivalent circular radius. Of course, this radius is less than that assuming a spherical shape.

Based on this approach, the permeation of the alcohol solutes decreases with increasing solute molecule radius, regardless of which way the radius was calculated, consistent with the decrease in the solute-accessible volume with probe radius, as shown in Fig. 13. However, Fig. 13 also demonstrates the difference between urea and the alcohols. The urea permeability through the membrane is lower than that of ethanol, even though the urea molecule is smaller in size. This is likely due to effects associated with more complex chemistry of urea in that urea has more hydrogen bonding sites than ethanol. This could make it harder for urea to shed hydrogen bonded water molecules as it passes into and through the membrane. It may also result in hydrogen bonds between the urea and functional groups of the membrane polymer chains. Further work is necessary to elucidate this. The rejections of Na+ and Cl- are both larger than all organic solutes, even though their dehydrated solute volumes are about the same as methanol. This is likely due to the difference between the nature of ion-water chemical interaction and organic solute-water chemical interaction. The effects of chemistry on solute transport and rejection are the subject of our current research efforts.

## 4. Conclusions

The Non-Equilibrium Molecular Dynamics (NEMD) methodology that we used here provides a powerful approach to studying water purification in polymeric RO membranes. Instead of focusing on macroscale transport models, NEMD allows the direct simulation of solvent and solute molecules interacting with each other and the polymeric chains of the membrane. The challenge with these simulations is two-fold. First, we have to be certain that the molecular-level models for interactions between molecules are reasonably accurate. In this case, we can be fairly confident of this given the similarity between several aspects of the simulated membrane and actual physical membranes including the dry membrane density, hydrated membrane density, pure water flux, and pore sizes as well as the similar qualitative results for solute permeability and rejection. Second, we have to overcome the immense computational requirements of such simulations. The time scales are so short and the length scales so small that even the smallest simulation domains and shortest durations require significant computational resources. Here we are only able to model a membrane that is tens of Ångströms in size for only tens of nanoseconds with thousands of water molecules and hundreds of solute molecules. More comprehensive conclusions will require the exploration of more types of solutes, larger simulation systems, and longer simulation times. Nevertheless, in spite of these limitations, substantial understanding is gained—and more is possible.

In this case, we were able to investigate water, organic solute, and salt transport through the commonly used FT-30 polymeric reverse osmosis membrane. We find a strong correlation between water transport and the percolated free volume within the membrane. The solute rejection is positively correlated with the Van der Waals size of the dehydrated solutes for alcohol solutes. However, the dependence of solute rejection on the dehydrated size of the solutes breaks down for urea and ions. Thus, in order to accurately predict the solute rejection by RO membranes, both the size effect in terms of solute size and membrane structure and chemistry effects in terms of solute-water and solute-membrane interaction needs to be taken into account. Work is continuing on this aspect of the problem.

## 5. Acknowledgement

M. S. thanks Dr. Steve Arturo from Dow Company and Dr. Lev Sarkisov from Edinburgh University for useful discussions, as well as Northwestern University High Performance Computing Center for a supercomputing grant. Part of this work also used the Extreme Science and Engineering Discovery Environment (XSEDE) [65], which is supported by National Science Foundation grant number ACI-1053575. The authors gratefully acknowledge funding from the Institute for Sustainability and Energy at Northwestern (ISEN).

**Table captions:**

Table 1. The density and thickness of dry and hydrated membranes.
Table 2. The side and head-on views of space-filling models of four organic solutes and a water molecule drawn by VMD [42]. The area of the head-on view shows the minimum cross-section area.

**Figure captions:**

Fig. 1. (a) The monomers and the crosslinked amide, where green represents N, yellow represents Cl, blue represents C, white represents H, and red represents O, and (b) The membrane structure after the crosslinking process, where colors represent different crosslinked chains. (Color online.)

Fig. 2. Simulation setup for non-equilibrium molecular dynamics (NEMD) simulations with membrane M1. Water molecules are red, 2-propanol molecules in the solution to the left of the membrane are blue, graphene carbon atoms are green, and membrane atoms are gray except those that are pinned to a fixed position in space, which are yellow. (Color online).

Fig. 3. The number of water molecules transported through the membrane as a function of time for pure water at various pressures for membrane M1.

Fig. 4. Water molecule flux as a function of pressure for (■) pure water, (○) methanol, (▲) ethanol, (●) 2-propanol, (▼) urea, and (□) NaCl solutions on the left side of the membrane. Dashed lines represent extrapolations based on the theoretical osmotic pressure. (Color online.)

Fig. 5. The number of water molecules transported through the membrane as a function of time in membrane M1 for pure water and methanol, ethanol, 2-propanol, urea, and NaCl solutions at 150 MPa. (Color online.)

Fig. 6. The free volume distribution for hydrated membranes (a) M1, (b) M2, (c) M3 and (d) M4. Each color represents a 0.5 Å thick plane at a different depth in the x direction. (Color online.)

Fig. 7. The permeability coefficient $K$ (solid squares) as a function of the percolated free volume percentage in the hydrated state for four different membranes.

Fig. 8. (a) The percolated water-accessible free volume accumulated over 10 ns in the dense region of the membrane M1 and (b) the trajectories of water molecules that passed through the membrane M1 over 10 ns for 2 Å thick slices at x = -20 Å, -10 Å, 0 Å, 10 Å and 20 Å. The color represents the depth in the x direction. (Color online.)

Fig. 9. The free volume for membrane M1 (a) in the dry state, (b) at a single frame in the hydrated state, and (c) accumulated over 10 ns in the hydrated state. (d) The percentage of free volume in the dense membrane region for the dry membrane M1, hydrated membrane M1 at a single frame and hydrated membrane accumulated over 10 ns. (Color online.)

Fig. 10. Pore size distribution for the dry membrane M1 after the construction processes (dashed curve) and for the summation of over 10 ns in hydrated states in NEMD simulations (solid curve).

Fig. 11. Trajectories at a transmembrane pressure of 150 MPa for (a, b) methanol, (c, d) ethanol, (e, f) 2-propanol and (g, h) urea molecules illustrating "hopping" mechanisms, especially clear for 2-propanol. Residence time is indicated for selected pores and paths between pores. The approximate location of the membrane surface is indicated by the vertical dashed lines.

Fig. 12. The solute number density (a,c,e,g,i) before applying a reverse osmotic pressure and (b,d,f,h,j) 20 ns after applying 150 MPa pressure to the graphene layer on the left. The gray region represents the membrane.

Fig. 13. The number of solute molecules passing through the central plane of the M1 membrane when 1000 water molecules have permeated through the membrane as a function of solute radius estimated from the solute volumes assuming a spherical shape (solid squares) and estimated from the minimum cross-section of the solutes assuming a circular cross-section (open circles). Both the solute volume and area are calculated from a grid-counting algorithm. The solute-accessible volume, or free volume, within the membrane is a function of the probe radius (solid triangles).

Supplementary materials

V1. Video of 2-propanol transport within the RO membrane molecular structure (50.7 ns duration).

V2. Video of the trajectory of a 2-propanol molecule (38.9 ns duration).

F1. Supplementary figures.

[1] WHO/UNICEF, Progress on drinking water and sanitation, in: Joint Monitoring Programme Report 2014, 2014.
[2] R.W. Baker, Membrane technology and applications, 2nd ed., John Wiley & Sons, Ltd., Chichester, 2004.
[3] K.P. Lee, T.C. Arnot, D. Mattia, A review of reverse osmosis membrane materials for desalination—Development to date and future potential, J. Membr. Sci., 370 (2011) 1-22.
[4] Y. Kiso, Y. Sugiura, T. Kitao, K. Nishimura, Effects of hydrophobicity and molecular size on rejection of aromatic pesticides with nanofiltration membranes, J. Membr. Sci., 192 (2001) 1-10.
[5] Y. Yoon, R. Lueptow, Removal of organic contaminants by RO and NF membranes, J. Membr. Sci., 261 (2005) 76-86.
[6] L.F. Greenlee, D.F. Lawler, B.D. Freeman, B. Marrot, P. Moulin, Reverse osmosis desalination: water sources, technology, and today's challenges, Water Res., 43 (2009) 2317-2348.
[7] S.H. Kim, S.Y. Kwak, T. Suzuki, Positron annihilation spectroscopic evidence to demonstrate the flux-enhancement mechanism in morphology-controlled thin-film-composite (TFC) membrane, Environ. Sci. Technol. , 39 (2005) 1764-1770.
[8] J.G. Wijmans, R.W. Baker, The solution diffusion model: a review, J. Membr. Sci., 107 (1995) 1-21.
[9] H. Yasuda, A. Peterlin, Diffusive and bulk flow transport in polymers, J. Appl. Polymer Sci., 17 (1973) 433-442.
[10] W.R. Bowen, A.W. Mohammad, Characterization and prediction of nanofiltration membrane performance—a general assessment, Trans. Inst. Chem. Eng., 76 (1998) 885-893.
[11] S. Lee, R.M. Lueptow, Membrane rejection of nitrogen compounds, J. Membr. Sci., 35 (2001) 3008-3018.
[12] F. Peng, Z. Jiang, E.M.V. Hoek, Tuning the molecular structure, separation performance and interfacial properties of poly(vinyl alcohol)–polysulfone interfacial composite membranes, J. Membr. Sci., 368 (2011) 26-33.
[13] M.P. Allen, Computational soft matter: from synthetic polymers to proteins lecture notes, Introduction to Molecular Dynamics Simulation, John von Neumann Institute for Computing, Jülich, Germany, 2004.
[14] L.Y. Wang, R.S. Dumont, J.M. Dickson, Nonequilibrium molecular dynamics simulation of water transport through carbon nanotube membranes at low pressure, J. Chem. Phys., 137 (2012) 044102.
[15] A. Kalra, S. Garde, G. Hummer, Osmotic water transport through carbon nanotube membranes, Proc. Natl. Acad. Sci., 100 (2003) 10175-10180.

[16] G. Hummer, J.C. Rasaiah, J.P. Noworyta, water conduction through the hydrophobic channel of a carbon nanotube, Nature, 414 (2001) 188 – 190.
[17] A. Berezhkovskii, G. Hummer, Single-file transport of water molecules through a carbon nanotube, Phys. Rev. Lett., 89 (2002) 064503.
[18] Z. Hu, Y. Chen, J. Jiang, Zeolitic imidazolate framework-8 as a reverse osmosis membrane for water desalination: insight from molecular simulation, J. Chem. Phys., 134 (2011) 134705.
[19] T. Yoshioka, M. Asaeda, T. Tsuru, A molecular dynamics simulation of pressure-driven gas permeation in a micropore potential field on silica membranes, Journal of Membrane Science, 293 (2007) 81-93.
[20] H. Frentrup, K.E. Hart, C.M. Colina, E.A. Muller, In Silico Determination of Gas Permeabilities by Non-Equilibrium Molecular Dynamics: $CO_2$ and He through PIM-1, Membranes (Basel), 5 (2015) 99-119.
[21] W.J. Koros, G.K. Fleming, Membrane-based gas separation, J. Membr. Sci., 83 (1993) 1-80.
[22] S.-T. Hwang, Fundamentals of membrane transport, Korean Journal of Chemical Engineering, 28 (2010) 1-15.
[23] L.A. Richards, A.I. Schafer, B.S. Richards, B. Corry, The importance of dehydration in determining ion transport in narrow pores, Small, 8 (2012) 1701-1709.
[24] H. Ebro, Y.M. Kim, J.H. Kim, Molecular dynamics simulations in membrane-based water treatment processes: A systematic overview, J. Membr. Sci., 438 (2013) 112-125.
[25] J. Zheng, E.M. Lennon, H.K. Tsao, Y.J. Sheng, S. Jiang, Transport of a liquid water and methanol mixture through carbon nanotubes under a chemical potential gradient, J. Chem. Phys., 122 (2005) 214702.
[26] F.Q. Zhu, K. Schulten, Water and proton conduction through carbon nanotubes as models for biological channels, Biophys. J., 85 (2003) 236-244.
[27] M.E. Suk, A.V. Raghunathan, N.R. Aluru, Fast reverse osmosis using boron nitride and carbon nanotubes, Appl. Phys. Lett. , 92 (2008) 133120.
[28] J.E. Cadotte, R.S. King, R.J. Majerle, R.J. Petersen, Interfacial synthesis in the preparation of reverse osmosis membranes, J. Macromol. Sci. Part A: Chem., 15 (1981) 727-755.
[29] M.J. Kotelyanskii, N.J. Wagner, M.E. Paulaitis, Molecular dynamics simulation study of the mechanisms of water diffusion in a hydrated, amorphous polyamide, Comp. Theor. Poly. Sci., 9 (1999) 301-306.
[30] M.J. Kotelyanskii, N.J. Wagner, M.E. Paulaitis, Atomistic simulation of water and salt transport in the reverse osmosis membrane FT-30, J. Membr. Sci., 139 (1998) 1-16.
[31] E. Harder, D.E. Walters, Y.D. Bodnar, R.S. Faibish, B. Roux, Molecular dynamics study of a polymeric reverse osmosis membrane, J. Phys. Chem. B, 113 (2009) 10177-10182.
[32] Y. Luo, E. Harder, R.S. Faibish, B. Roux, Computer simulations of water flux and salt permeability of the reverse osmosis FT-30 aromatic polyamide membrane, J. Membr. Sci., 384 (2011) 1-9.
[33] Z.E. Hughes, J.D. Gale, A computational investigation of the properties of a reverse osmosis membrane, J. Mater. Chem., 20 (2010) 7788.
[34] V. Kolev, V. Freger, Hydration, porosity and water dynamics in the polyamide layer of reverse osmosis membranes: A molecular dynamics study, Polym., 55 (2014) 1420-1426.
[35] L.D. Nghiem, A.I. Schafer, M. Elimelech, Removal of natural hormones by nanofiltration membranes: measurement, modeling, and mechanisms, Environ. Sci. Technol., 38 (2004) 1888-1896.
[36] F.A. Pacheco, I. Pinnau, M. Reinhard, J.O. Leckie, Characterization of isolated polyamide thin films of RO and NF membranes using novel TEM techniques, J. Membr. Sci., 358 (2010) 51-59.
[37] Y. Wang, P. Keblinski, Effect of interfacial interactions and nanoscale confinement on octane melting, Journal of Applied Physics, 111 (2012) 064321.
[38] J. Wang, R.M. Wolf, J.W. Caldwel, P.A. Kollman, D.A. Case, Development and testing of a general amber force field, J. Comput. Chem., 25 (2004) 1157–1174.
[39] J.C. Phillips, R. Braun, W. Wang, J. Gumbart, E. Tajkhorshid, E. Villa, C. Chipot, R.D. Skeel, L. Kale, K. Schulten, Scalable molecular dynamics with NAMD, J. Comput. Chem., 26 (2005) 1781-1802.
[40] M.P. Allen, D.J. Tildesley, Computer simulations of liquids, Oxford University Press, Inc., New York, NY, 1987.

[41] T. Darden, D. York, L. Pedersen, Particle mesh Ewald: An N·log(N) method for Ewald sums in large systems, J. Chem. Phys., 98 (1993) 10089.
[42] W. Humphrey, A. Dalke, K. Schulten, VMD: Visual molecular dynamics, J. Mol. Graphics, 14 (1996) 33-38.
[43] S. Shenogin, R. Ozisik, Xenoview: visualization for atomistic simulations, in: http://xenoview.mat.rpi.edu, 2009.
[44] G.A. Ozpinar, W. Peukert, T. Clark, An improved generalized AMBER force field (GAFF) for urea, J. Mol. Model., 16 (2010) 1427-1440.
[45] J. Wang, W. Wang, P.A. Kollman, D.A. Case, Automatic atom type and bond type perception in molecular mechanical calculations, J. Mol. Graph. Model., 25 (2006) 247-260.
[46] A. Jakalian, D.B. Jack, C.I. Bayly, Fast, efficient generation of high-quality atomic charges. AM1-BCC model: II. Parameterization and validation, J. Comput. Chem., 23 (2002) 1623-1641.
[47] D. Beglov, B. Roux, Finite representation of an infinite bulk system: Solvent boundary potential for computer simulations, J. Chem. Phys., 100 (1994) 9050.
[48] W.L. Jorgensen, J. Chandrasekhar, J.D. Madura, R.W. Impey, M.L. Klein, Comparison of simple potential functions for simulating liquid water, J. Chem. Phys., 79 (1983) 926.
[49] S. Gumma, Gibbs dividing surface and helium adsorption, Adsorp., 9 (2003) 17-28.
[50] X. Zhang, D.G. Cahill, O. Coronell, B.J. Mariñas, Absorption of water in the active layer of reverse osmosis membranes, J. Membr. Sci., 331 (2009) 143-151.
[51] B. Mi, D.G. Cahill, B.J. Mariñas, Physico-chemical integrity of nanofiltration/reverse osmosis membranes during characterization by Rutherford backscattering spectrometry, J. Membr. Sci., 291 (2007) 77-85.
[52] R.I. Urama, B.J. Marifias, Mechanistic interpretation of solute permeation through a fully aromatic polyamide reverse osmosis membrane, J. Membr. Sci., 123 (1997) 267-280.
[53] R.A. Robinson, R.H. Stokes, Electrolyte Solutions, Dover Publications, Inc, Mineola, NY, 2002.
[54] Y. Luo, B. Roux, Simulation of osmotic pressure in concentrated aqueous salt solutions, J. Phys. Chem. Lett., 1 (2010) 183-189.
[55] F. Franks, Water: A matrix of life, (Second Edition), Royal Society of Chemistry, Cambridge, 2000.
[56] L. Sarkisov, A. Harrison, Computational structure characterisation tools in application to ordered and disordered porous materials, Mol. Simul., 37 (2011) 1248-1257.
[57] J. Hoshen, R. Kopelman, Percolation and cluster distribution. I. Cluster multiple labeling technique and critical concentration algorithm, Phys. Rev. B, 14 (1976) 3438-3445.
[58] S. Turgman-Cohen, J.C. Araque, E.M. Hoek, F.A. Escobedo, Molecular dynamics of equilibrium and pressure-driven transport properties of water through LTA-type zeolites, Langmuir, 29 (2013) 12389-12399.
[59] M.T.M. Pendergast, E.M. Hoek, A review of water treatment membrane nanotechnologies, Environ. Sci. Technol., 4 (2011) 1946-1971.
[60] L. Ruiz, Y. Wu, S. Keten, Tailoring the water structure and transport in nanotubes with tunable interiors, Nanoscale, 7 (2014) 121-132.
[61] L.B. Pártay, P. Jedlovszky, Á. Vincze, G. Horvai, Properties of free surface of water−methanol mixtures. Analysis of the truly interfacial molecular layer in computer simulation, J. Phys. Chem. B, 112 (2008) 5428-5438.
[62] L.R. Pratt, A. Pohorille, Hydrophobic effects and modeling of biophysical aqueous solution interfaces, Chem. Rev., 102 (2002) 2671-2692.
[63] D. Cohen-Tanugi, J.C. Grossman, Water desalination across nanoporous graphene, Nano Lett., 12 (2012) 3602-3608.
[64] D. Bemporad, C. Luttmann, J.W. Essex, Computer simulation of small molecule permeation across a lipid bilayer: dependence on bilayer properties and solute volume, size, and cross-sectional area, Biophys. J., 87 (2004) 1-13.

[65] J. Towns, T. Cockerill, M. Dahan, I. Foster, K. Gaither, A. Grimshaw, V. Hazlewood, S. Lathrop, D. Lifka, G.D. Peterson, R. Roskies, J.R. Scott, N. Wilkins-Diehr, XSEDE: Accelerating Scientific Discovery, Comp. Sci. Eng., 16 (2014) 62-74.